\documentclass[%
 reprint,
superscriptaddress,
showpacs,
nofootinbib,
 amsmath,amssymb,aps,
prd,
floatfix,
]{revtex4-1}

\usepackage{lipsum} 

\usepackage{bm}
\usepackage{comment}
\pdfoutput=1
\usepackage{amsmath,amsfonts,amsthm}
\usepackage{esdiff}  
\usepackage{booktabs}  
\usepackage{url}  
\usepackage[hidelinks]{hyperref}  
\usepackage{relsize}

\usepackage{cleveref}  
	\crefname{equation}{equation}{equations}
	\crefname{figure}{figure}{figures}	
	\crefname{table}{table}{tables}
\usepackage[caption=false]{subfig}

\usepackage[normalem]{ulem}

\usepackage[usenames,dvipsnames,svgnames,table]{xcolor}

\usepackage[export]{adjustbox}
\usepackage{graphicx}

\usepackage[sc]{mathpazo} 
\usepackage[T1]{fontenc} 
\usepackage{microtype} 

\usepackage{booktabs} 
\usepackage{float} 
\usepackage{hyperref} 

\usepackage{lettrine} 
\usepackage{paralist} 

\usepackage{titlesec} 
\renewcommand\thesection{\Roman{section}} 
\renewcommand\thesubsection{\Alph{subsection}} 
\titleformat{\section}[block]{\large\scshape\centering\bfseries}{\thesection.}{1em}{} 

\titleformat{\subsection}[block]{\scshape\centering}{\thesubsection.}{1em}{} 

\DeclareCaptionFormat{myformat}{#1#2#3\hrulefill}
\usepackage{float}
\floatstyle{plaintop}
\restylefloat{table}

\DeclareMathOperator{\Tr}{Tr}

\begin{document}

\title{Inference of bipolar neutrino flavor oscillations near a core-collapse supernova, based on multiple measurements at Earth} 

\author{Eve Armstrong}
\email{evearmstrong.physics@gmail.com}
\affiliation{Department of Physics, New York Institute of Technology, New York, NY 10023, USA}
\affiliation{Department of Astrophysics, American Museum of Natural History, New York, NY 10024, USA}
\author{Amol V.\ Patwardhan}
\email{apatward@slac.stanford.edu} 
\affiliation{SLAC National Accelerator Laboratory, Menlo Park, CA 94025, USA}
\author{A.A. Ahmetaj}
\affiliation{Department of Physics, New York Institute of Technology, New York, NY 10023, USA}
\author{M. Margarette Sanchez}
\affiliation{Department of Physics, New York Institute of Technology, New York, NY 10023, USA}
\author{Sophia Miskiewicz}
\affiliation{Department of Physics, Stevens Institute of Technology, Hoboken, NJ 07030, USA}
\author{Marcus Ibrahim}
\affiliation{Department of Physics, New York Institute of Technology, New York, NY 10023, USA}
\author{Ishaan Singh}
\affiliation{Department of Physics, New York Institute of Technology, New York, NY 10023, USA}
\date{\today}

\begin{abstract}
{Neutrinos in compact-object environments, such as core-collapse supernovae, can experience various kinds of collective effects in flavor space, engendered by neutrino-neutrino interactions. These include \lq\lq bipolar\rq\rq\ collective oscillations, which are exhibited by neutrino ensembles where different flavors dominate at different energies. Considering the importance of neutrinos in the dynamics and nucleosynthesis in these environments, it is desirable to ascertain whether an Earth-based detection could contain signatures of bipolar oscillations that occurred within a supernova envelope. To that end, we continue examining a cost-function formulation of statistical data assimilation (SDA) to infer solutions to a small-scale model of neutrino flavor transformation.  SDA is an inference paradigm designed to optimize a model with sparse data.  Our model consists of two mono-energetic neutrino beams with different energies emanating from a source and coherently interacting with each other and with a matter background, with time-varying interaction strengths.  We attempt to infer flavor transformation histories of these beams using simulated measurements of the flavor content at locations \lq\lq in vacuum\rq\rq\ (that is, far from the source), which could in principle correspond to earth-based detectors. Within the scope of this small-scale model, we found that: (i) based on such measurements, the SDA procedure is able to infer \textit{whether} bipolar oscillations had occurred within the protoneutron star envelope, and (ii) if the measurements are able to sample the full amplitude of the neutrino oscillations in vacuum, then the amplitude of the prior bipolar oscillations is also well predicted. This result intimates that the inference paradigm can well complement numerical integration codes, via its ability to infer flavor evolution at physically inaccessible locations.}
\end{abstract}

\maketitle 
\maketitle 

\section{Introduction} \label{sec:intro}

The physics of neutrino flavor evolution can significantly influence the dynamics and nucleosynthesis in core-collapse supernovae (CCSN) and neutron star binary mergers~\cite{Fuller:1992eu,Qian:1993dg,Fuller:1993ry,Fuller:1995qy,Duan:2010af,Wu:2014kaa,Wu:2015glr,Sasaki:2017jry,Balantekin:2017bau,Xiong:2019nvw,Xiong:2020ntn}, and the era of multi-messenger astrophysics offers us an unprecedented vantage point on these events. Understanding flavor evolution is critical for leveraging gravitational wave and electromagnetic observations so as to deepen our understanding of energy, entropy, and lepton number transport at these sites. Owing to the large fluxes of neutrinos in these environments, their flavor evolution can be significantly impacted by neutrino coherent forward scattering off of other neutrinos, resulting in a variety of interesting collective phenomena in flavor space (see, e.g., the reviews in \cite{Duan:2009cd,Duan:2010bg,Mirizzi:2015eza,Chakraborty:2016yeg,Tamborra:2020cul} and references therein).

One important question pertaining to neutrino flavor evolution in a CCSN environment is to ascertain whether \lq\lq bipolar\rq\rq\ oscillations~\cite{Kostelecky:1994dt,Samuel:1995ri,Duan:2005cp,Hannestad:2006nj,Duan:2006an,Duan:2006jv,Raffelt:2007yz} occur within the supernova envelope. These are driven by neutrino-neutrino coherent forward scattering, and arise in systems where the initial state of the interacting neutrino ensemble exhibits a dominance of different flavors at different energies. Bipolar oscillations involve neutrinos at different energies rapidly and repeatedly swapping flavors as they propagate. In numerical solutions of flavor evolution in CCSN environments, neutrinos typically experience these types of oscillations at earlier radii; that is, prior to undergoing the Mikheyev-Smirnov-Wolfenstein (MSW) resonance~\cite{mikheev2007neutrino,mikheev1985resonance,wolfenstein1978neutrino}, a phenomenon that arises due to coherent forward scattering with matter.  These oscillations dramatically change the flavor evolution histories of neutrinos compared to the otherwise-simple MSW-only scenario. As a result of occurring deeper within the supernova envelope (where the neutrino fluxes are higher) compared to the MSW resonance, the effects of bipolar oscillations are potentially more significant with regard to energy transport and nucleosynthesis.

There exist powerful numerical integration codes for obtaining solutions to the flavor evolution problem in compact object environments~\cite{duan2006simulation,duan2008simulating,richers2019neutrino}. Utilizing these codes, however, requires making definite choices regarding the relative flavor content of the neutrinos at the point of emission from the proto-neutron star, based on reasonable assumptions about the the physics of dense nuclear matter and neutrino decoupling. Typically, the decoupling of neutrinos from chemical and thermal equilibrium is approximated to be instantaneous at the surface of the proto-neturon star, represented by a single, sharp \lq\lq neutrino-sphere\rq\rq. As a result, the initial states of neutrinos at the neutrino-sphere radius are taken to be definite flavor states.  It has been shown in recent years that relaxing these assumptions regarding uniform, instantaneous neutrino decoupling can result in an emission-angle dependence in the initial flavor content of neutrinos, resulting in the phenomenon of \lq\lq fast\rq\rq\ flavor oscillations (Ref.~\cite{Tamborra:2020cul} and references therein). 

Moreover, including the effects of direction-changing scattering of neutrinos can result in a small, non-outward-propagating component of the neutrino flux, which can nevertheless significantly contribute to the forward-scattering potential experienced by the outgoing neutrinos, as a result of the large intersection angles between their trajectories~\cite{cherry2012neutrino,Cherry:2013mv,Zaizen:2019ufj,Cherry:2019vkv}. This \lq\lq halo effect\rq\rq\ potentially changes how this problem must be approached --- not as an initial-value problem with flavor content fully specified at the source surface, but rather as a boundary-value problem, with flavor information propagating both outward and inward. As a result of such discoveries, it has become pertinent to ask how much can be learned about neutrino oscillations near a supernova from a future earth-based neutrino detection, without any \textit{a priori} assumptions about the initial conditions~\cite{rrapaj2021inference}.

In this paper, we avoid assumed knowledge of flavor evolution at inaccessible locations within the supernova envelope, by adopting an inverse approach.  Using a small-scale model with simulated data, we ask: what information can we \textit{infer} regarding the realm of bipolar oscillations, using measurements made only in the vacuum oscillations regime?  

Specifically, we seek to ascertain whether multiple measurements of flavor made in vacuum contain a signature of the flavor evolution history within the supernova envelope, where the neutrino-matter and neutrino-neutrino potentials are dominant. By \lq\lq multiple\rq\rq\ measurements, we mean: measurements spaced out in location but clustered within the vacuum oscillations regime --- a proxy for multiple Earth-based detectors. Importantly, the critical differences between this formulation and the forward integration approach are that we do not assume knowledge of (i) unmeasurable model state variables, or (ii) any (measurable or unmeasurable) state variables at physically inaccessible locations.  We ask whether the accessible information is sufficient to infer the complete flavor transformation histories of neutrinos back to the emission surface.

To adopt this formalism we employ an inference procedure.  Inference is a means to optimize a model given measurements, where the measurements are assumed to arise from model dynamics.  Importantly, an inference procedure need not be formulated as an initial-value problem.  Rather, we formulate the procedure using partial information at one bound (near Earth) and zero information at the other (at emission).

The specific inference technique used in this paper is statistical data assimilation (SDA).  SDA was invented for numerical weather prediction~\cite{kimura2002numerical,kalnay2003atmospheric,evensen2009data,betts2010practical,whartenby2013number,an2017estimating} for the case of sparse data.  It has since gained traction in neurobiology~\cite{schiff2009kalman,toth2011dynamical,kostuk2012dynamical,hamilton2013real,meliza2014estimating,nogaret2016automatic,armstrong2020statistical}, for estimating cellular and synaptic properties given sparse neuronal electrical signals. Within astrophysics, the known applications of SDA include exoplanet modelling~\cite{Madhusudhan2018} and solar cycle prediction~\cite{Kitiashvili2008,kitiashvili2020}. In recent years, the utility of SDA has been explored in the context of inferring solutions to small-scale flavor evolution models~\cite{armstrong2017optimization,armstrong2020inference,rrapaj2021inference,armstrong2021inference}.

In this paper we find, for a small-scale steady-state coherent forward-scattering model of flavor evolution, that multiple measurements of neutrino flavor in the vacuum-oscillations regime could contain a signature of the frequency and amplitude --- and to some degree the complexity of the waveform --- of bipolar oscillations that had occurred near the point of emission.  We quantify the robustness of this result, and discuss implications regarding a real detection. 

\section{Model} \label{sec:model} 

\subsection{\textbf{Formulation}} \label{sec:modelForm}
Our model has been fully described in Refs.~\cite{armstrong2017optimization,armstrong2020inference,rrapaj2021inference}, and we refer the reader there for details.  Here we briefly describe the model's equations of motion, and note one important feature of the collective neutrino oscillations problems: nonlinearity.  

We consider a single-angle, two-flavor scenario wherein two mono-energetic neutrino beams with different energies interact with each other and with a background consisting of particles carrying weak charge, such as nuclei, free nucleons, and electrons.  The densities of the background particles and of the neutrino beams dilute as some functions of a position coordinate $r$, which we interpret as the distance from the neutrino-sphere in a supernova.  That is: on their journey through the supernova envelope, the neutrinos interact coherently with each other and with the dense ejecta surrounding the star immediately after core collapse.  Importantly, the model is a forward-scattering-only scenario, rendering it solvable via traditional forward-integration techniques --- a consistency check for SDA solutions.

We write the equations of motion for flavor evolution of each neutrino in terms of \lq\lq polarization vectors\rq\rq\ $\vec P_i$, after decomposing the density matrices and Hamiltonians, respectively, into bases of Pauli spin matrices\footnote{The polarization vectors, or Bloch vectors, are defined in terms of the neutrino density matrices: $\rho_i = \frac12(\mathbb{1}+\vec\sigma \cdot \vec P_i$). The Hamiltonian can be decomposed in the same manner as $H_i = \frac12(\Tr(H_i) + \vec\sigma \cdot \vec V_i)$.  Here, $\vec V_i$ contains contributions from vacuum oscillations, neutrino-matter interactions, and neutrino-neutrino interactions, as shown in Eq.~\ref{eq:model}.} (for details see Ref.~\cite{Raffelt1993,Sigl1993}):
\begin{equation} \label{eq:model}
  \diff{\vec{P}_{i}}{r} = \left(\Delta_i \vec{B} + V(r) \hat{z} 
  +\mu(r) \sum_{j\neq i} \vec{P}_j \right) \times \vec{P}_i
\end{equation}

In Equation~\ref{eq:model}, $\Delta_{i} = \delta m^2/(2E_{i})$ are the vacuum oscillation frequencies of the two neutrinos with energies $E_1$ and $E_2$, with $\delta m^2$ being the mass-squared difference in vacuum.  The unit vector $\vec{B}=\sin(2 \theta) \hat{x} -\cos(2 \theta) \hat{z}$ represents flavor mixing in vacuum, with mixing angle $\theta$.  The functions $V(r)$ and $\mu(r)$ are potentials for neutrino-matter and neutrino-neutrino coupling, respectively.  They take the forms $V(r) = C_m/(r+r_m)^3$ and $\mu(r) = Q/(r+r_\nu)^4$, respectively; $C_m$ and $Q$ are  constant numbers, and $r_m$ and $r_\nu$ are offsets which determine the reference values of $V$ and $\mu$ at $r=0$. This form for the neutrino-neutrino coupling reflects the manner in which coupling strength varies in the neutrino bulb model calculations that employ the single-angle approximation.  All model parameters are taken to be constant and known to the SDA procedure (Table~\ref{table:Known}) throughout all of the experimental setups described in this work, with the exception of Sec.~\ref{sec:resultsParamEst}.  We again emphasize that the equations of motion are fiercely nonlinear --- and that SDA was designed to perform state-and-parameter estimation for nonlinear models.

The $\hat{z}$ component of the neutrino polarization vector denotes the net flavor content of electron flavor minus \lq\lq $x$\rq\rq\ flavor, the latter being a superposition of muon and tau flavors.  In this scenario, we assume that flavor evolution is driven entirely by coherent forward-scattering.  At certain distances from the emission surface for each neutrino, the forward scattering potential arising from neutrino-matter and neutrino-neutrino interactions leads to an in-medium effective neutrino mass level crossing, referred to as the \lq\lq MSW resonance.\rq\rq~\cite{mikheev2007neutrino,mikheev1985resonance,wolfenstein1978neutrino}. The MSW resonance is associated with an enhanced $e$ $\leftrightarrow$ $x$ flavor conversion probability.

\setlength{\tabcolsep}{5pt}
\begin{table}[H]
\small
\centering
\begin{tabular}{|l |c | l | c |} \toprule
\hline
 \textit{Parameter} & \textit{Value} &  \textit{Parameter} & \textit{Value} \\\midrule \hline
 $\Delta_1$ & 1000 & $\Delta_2$ & 2500 \\
 Q & 100.0 & C & 3308 \\
 $r_\nu$ & 0.51 & $r_m$ & 0.50 \\
 $\theta$ & 0.1 & & \\\bottomrule \hline
\end{tabular}
\caption{\textbf{Model parameters taken to be known and fixed during the estimation procedure.}  The $\Delta_i$ are the vacuum oscillation frequencies of the neutrinos, and $C$ and $Q$ are the multiplicative factors governing the neutrino-matter, and neutrino-neutrino coupling potentials $V(r)$ and $\mu(r)$, respectively, and $r_m$ and $r_\nu$ are the radial offsets. Parameter $\theta$ is the mixing angle in vacuum. Each of these parameters are taken to be known in all of the SDA experiments, with the exception of the matter coefficient $C$ being left unknown in one experiment (Section~\ref{sec:resultsParamEst}), where the SDA procedure was tasked with inferring its value.} 
\label{table:Known}
\end{table}

\subsection{Bipolar oscillations} \label{sec:modelBipolar}

Collective neutrino oscillations in spherically symmetric models are known to generically exhibit two types of flavor oscillation phenomena: \lq\lq synchronized\rq\rq\ and \lq\lq bipolar\rq\rq~\cite{Duan:2005cp,Hannestad:2006nj,Duan:2006an,Duan:2006jv,Duan:2010bg}. The synchronized mode is exhibited, for instance, in a dense neutrino gas where all the neutrinos are initially of the same flavor; that is: all the individual polarization vectors $P_\omega$ are aligned with one another. This can cause the system to collectively oscillate with a unified angular frequency $\Omega_\text{sync}$.

The bipolar mode, on the other hand, may be exhibited by systems consisting of Polarization vectors pointing in opposite directions (e.g., $\nu_e$ and $\nu_x$), or $\nu_e$ and $\bar\nu_e$. Bipolar oscillations can be understood analytically by considering a simple toy system of two neutrinos with equal and oppositely aligned polarization vectors.  This simple two-neutrino system permits an analogy with an inverted pendulum (the initial state for bipolar oscillations represents an unstable equilibrium configuration). Through this analogy, it can be shown that the characteristic frequency of these bipolar oscillations is $\sim \sqrt{\Delta \mu}$~\cite{Hannestad:2006nj}, where $\Delta$ and $\mu$ are the vacuum oscillation frequency and the neutrino-neutrino interaction strength, as defined in Eq.~\eqref{eq:model}.

\section{Method} \label{sec:method}

\subsection{General formulation} \label{sec:methodGen}

Statistical data assimilation is an inference procedure wherein any measured quantities are assumed to arise from the dynamics of a physical model, which may be nonlinear in nature, and where only a subset of the state variables can be experimentally accessed.  This model $\bm{F}$ can be written as a set of $D$ ordinary differential equations that evolve in some parameterization $r$ as:
\begin{align}
  \diff{x_a(r)}{r} &= F_a(\bm{x}(r),\bm{p}(r)); \hspace{1em} a =1,2,\ldots,D,
\end{align}
\noindent
where the components $x_a$ of the vector $\bm{x}$ are the model state variables.  Unknown parameters to be estimated are contained in $\bm{p}$, and may be variable.

A subset $L$ of the $D$ state variables is associated with measured quantities.  One seeks to estimate the evolution of all state variables that is consistent with the measurements provided, to predict model evolution at parameterized locations where measurements are not present.  

A prerequisite for estimation using real experimental data is the design of simulated experiments, where the true model evolution is known.  Simulated experiments offer the opportunity to ascertain which and how few experimental measurements and constraints, in principle, are sufficient to complete a model.  This is a critical question for cases wherein available measurements are extremely sparse --- as will be the case for an Earth-based neutrino detection from a future core-collapse supernova.  Finally, in this paper, we use forward integration to generated simulated data, as a consistency check for SDA solutions.  

\subsection{Optimization formulation} \label{sec:methodOpt}

We formulate the SDA procedure as an optimization wherein a cost function is extremized, and we write the cost function in two terms.  One term represents \lq\lq measurement error\rq\rq\:  the difference between state prediction and any measurements made.  The second term represents \lq\lq model error\rq\rq\: the difference between state prediction and adherence to the model dynamics\footnote{In previous works (Refs~\cite{armstrong2020inference,armstrong2017optimization}), the cost function also included an equality constraint to impose unitarity. Eliminating that term has two advantages. One is the easing of the computational burden. The other is that this makes the approach more amenable to flavor evolution studies including the collision terms~\cite{armstrong2021inference} --- a scenario in which unitarity is not necessarily conserved.}. It will be shown below in this Section that treating the model error as finite offers a systematic method to identify the lowest minimum, in a specific region of state-and-parameter space, of a non-convex cost function.  We search the surface of the cost function via the variational method.  The procedure in its entirety --- that is: a variational approach to minimization coupled with an annealing method to identify a lowest minimum of the cost function --- is referred to as variational annealing (VA). 

The cost function $A_0$ used in this paper is written as:
\begin{widetext}
\begin{equation} \label{eq:action}
\begin{split}
A_0 =& R_f A_{model} + R_m A_{meas}\\
A_{model}=&\frac{1}{{N}D}	\mathlarger{\sum}_{n \in \{\text{odd}\}}^{N-2} \mathlarger{\sum}_{a=1}^D \left[ \left\{x_a(n+2) - x_a(n) - \frac{\delta r}{6} [F_a(\bm{x}(n), \bm{p}) + 4F_a(\bm{x}(n+1),\bm{p}) + F_a(\bm{x}(n+2),\bm{p})]\right\}^2  \right. \\
  & \hspace{100pt} + \left.\left\{ x_a(n+1) - \frac12 \left(x_a(n)+x_a(n+2)\right) - \frac{\delta r}{8} [F_a(\bm{x}(n),\bm{p}) - F_a(\bm{x}(n+2),\bm{p})]\right\}^2 \right] \\
  A_{meas}=& \frac{1}{N_\text{meas}} \mathlarger{\sum}_j \mathlarger{\sum}_{l=1}^L (y_l(j) - x_l(j))^2.
\end{split}
\end{equation}
\end{widetext}
\noindent
One seeks the path $\bm{X}^0 = {\bm{x}(0),...,\bm{x}(N),\bm{p}(0),...\bm{p}(N)}$ in state space on which $A_0$ attains a minimum value.  One can derive this cost function by considering the classical physical Action on a path in a state space, where the path of lowest Action corresponds to the correct solution~\cite{abarbanel2013predicting}.  Hereafter we shall refer to the cost function of Eq.~\eqref{eq:action} as the Action. In a previous publication~\cite{armstrong2020inference}, it was shown that the action formulation offers a litmus test for identifying correct solutions: they are solutions that correspond to the path of least action.

The first squared term of Eq.~\eqref{eq:action} incorporates the model evolution of all $D$ state variables $x_a$. Here, the outer sum on $n$ is taken over all odd-numbered discretized radial locations of the model equations of motion.  The sum on $a$ is taken over all $D$ state variables\footnote{This term can be derived via consideration of Markov-chain transition probabilities~\cite{abarbanel2013predicting}.  For details, please also refer to Ref~\cite{armstrong2017optimization}.}.  In our model, the state variables are all three polarization components for each neutrino beam, or: $D=6$.

The second squared term of Equation~\ref{eq:action} governs the transfer of information from measurements $y_l$ to model states $x_l$.  Here, the summation on $j$ runs over all discretized radial locations $J$ at which measurements are made, which may be some subset of all integrated locations of the model.  The summation on $l$ is taken over all $L$ measured quantities\footnote{The measurement error term derives from the mutual information of probability theory~\cite{abarbanel2013predicting}.}.  In our model, these measured quantities are the $P_z$ component of the polarization vector for each neutrino beam, or: $L=2$.

The procedure searches a $(D \,(N+1)+ p)$-dimensional state space, where $N$ is the number of discretized steps, and $p$ is the number of unknown parameters in the model.  

\subsection{\textbf{Annealing to identify a lowest minimum of the cost function}} \label{sec:methodAnnealing}

Our model is nonlinear, and thus the Action surface will be non-convex.  The complete VA procedure anneals in terms of the ratio of model and measurement error, $R_f$ and $R_m$, respectively\footnote{More generally, $R_m$ and $R_f$ are inverse covariance matrices for the measurement and model errors, respectively.  In this paper the measurements are taken to be mutually independent, rendering these matrices diagonal.}, to gradually freeze out a lowest-minimum of the Action~\cite{ye2015systematic}.  This iteration works as follows.

We define the coefficient of measurement error $R_m$ to be 1.0, and write the coefficient of model error $R_f$ as: $R_f = R_{f,0}\alpha^{\beta}$, where $R_{f,0} = 10^{-1}$, $\alpha = 1.5$, and $\beta$ is initialized at zero.  Parameter $\beta$ is the annealing parameter.  When $\beta = 0$, relatively free from model constraints the Action surface is smooth and convex, and therefore there are no additional local minima.  Then we increase the weight of the model term slightly, via an integer increment in $\beta$, and recalculate the Action so that the procedure can again be tasked with finding the minimum.  We do this recursively toward the deterministic limit of $R_f \gg R_m$.  The aim is to remain sufficiently near to the lowest minimum so as not to become trapped in a local minimum as the surface acquires the structure imposed by the model dynamics.

\section{Experiments} \label{sec:expers}

\subsection{\textbf{Specific physics of interest: presence of bipolar oscillations?}} \label{sec:expersSpecs}

Using forward-integration simulations, we permit two neutrino beams of different energies to be emitted from the source (here the \lq\lq neutrino sphere\rq\rq\ of a proto-neutron star) in two different sets of flavor-state initial conditions.  In the first set, the two beams are emitted as pure electron-flavor eigenstates.  They evolve synchronously and smoothly through the MSW resonance.  In the second set, the beams are emitted in nearly opposite polarization states: one pure electron-flavor ($P_z = 1.0$) and the other nearly pure x-flavor ($P_z = -0.8$).  This second set gives rise to bipolar oscillations, as described in  Section~\ref{sec:modelBipolar}.  

We seek to examine these two scenarios for the following reason.  In a CCSN environment, typically the neutrino flux during the early shock breakout, or \lq\lq neutronization burst\rq\rq\ phase, is dominated by electron neutrinos over all other flavors of neutrinos and anti-neutrinos~\cite{muller2019neutrino}. Such initial conditions typically give rise to synchronous oscillations. Conversely, at later times during the supernova explosion, such as the neturino-driven wind phase, neutrinos are emitted in a rough equipartition among flavors, but with different average energies. As a result, there is a dominance of different flavors at different energies in the initial distribution of neutrinos --- leading to bipolar oscillations.  As the initial conditions will impact the subsequent flavor evolution and nucleosynthesis throughout the envelope, we seek to eliminate \textit{a priori} assumptions and instead ask what information regarding the early flavor evolution is contained in measurable quantities at an Earth-based detector. 

The challenge for the SDA procedure is to infer - based on measurements made in vacuum near Earth - which scenario had occurred at earlier radii: the synchronous behavior or the bipolar oscillations.  Translating to a larger-scale model, the question will become: \textit{In principle, can multiple measurements of flavor near Earth yield information about the flavor states at earlier radii within the matter-dominated region?}

\subsection{Details of the procedure} \label{sec:expersDetails}
\begin{figure*}[htb]
\centering
  \includegraphics[width=0.8\textwidth]{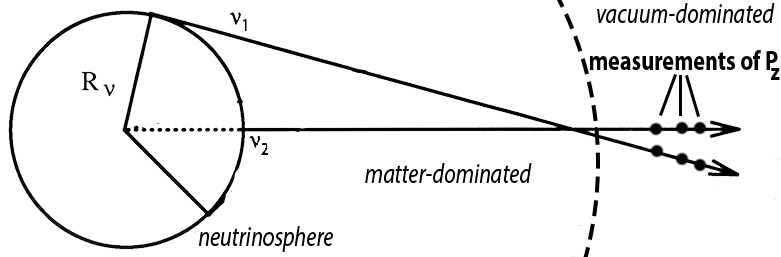}
  \caption{\textbf{Schematic of the small-scale simulation.}  Two neutrino beams, $\nu_1$ and $\nu_2$, emanate from an infinitely sharp "neutrino-sphere," which lies at radius $R_{\nu}$ from the center of the proto-neutron star.  The beams interact coherently through the matter-dominated envelope, and arrive at Earth.  Three detectors, clustered in the vacuum regime, sample the $P_z$ components (i.e. electron-flavor content) of each polarization vector.  In a realistic scenario, these three measurement locations represent three satellite- or Earth-based detectors.}
 \label{fig1}
\end{figure*}

In this paper, we give the SDA procedure full knowledge of the model parameters, and measurements were provided at three locations in the vacuum-dominated region.  Figure~\ref{fig1} offers a schematic.  Within the context of this simple model, by “measurement” we mean the value of $P_{1,z}$ and $P_{2,z}$ -- the z-component of the polarization vector for Neutrino Beams 1 and 2, respectively \footnote{Of course, a real detector will measure a spectrum convolved with contamination.}.  The procedure is provided no information regarding flavor outside of the three locations in vacuum.  The task is to take those sparse measurements, together with the model dynamics, to predict the complete flavor evolution history; that is: the values of $P_x$, $P_y$, and $P_z$ for each beam at each radial location between emission from the neutrino-sphere at $r=0$ and detection at $r \sim R$.  To obtain the prediction, the SDA procedure is permitted to search the full dynamical range for each variable -- of [-1.0:1.0] -- at each location\footnote{In previous publications (Refs~\cite{armstrong2020inference,armstrong2017optimization}), we used search ranges for $P_x$ and $P_y$ that were roughly three times stricter.  Broadening those ranges to encompass the full possible dynamics -- while it increased the computational expense -- has rendered the procedure more robust.}.  The confidence check on the SDA prediction is simulated "data" generated by forward-integration.

Our specific question is whether vacuum oscillations sampled near Earth contain a signature of whether bipolar oscillations occurred prior to the MSW resonance.  To this end, we perform two variations on the experimental design described above. 

In the first variation, the simulated data takes as initial conditions (at $r=0$) $P_{1,z}$ = 1.0 and $P_{2,z}$ = 1.0 ($P_x$ and $P_y$ are initialized at 0 for both neutrino beams.)  Initially aligned in the pure $\nu_e$ state, no bipolar oscillation occur, and the trajectory through the MSW resonance is smooth and synchronous.  The two beams emerge in nearly-pure $\nu_x$ flavor and no appreciable vacuum oscillations occur near the detector (at $r=R$).

In the second variation, the forward integration is instead initialized (at $r=0$) with $P_{1,z}=+1.0$ and $P_{2,z}=-0.8$ ($P_x$ and $P_y$ are again initialized at 0 for both beams.)  These represent two nearly opposite polarization states, which -- as described in Section~\ref{sec:modelBipolar} -- give rise to bipolar oscillations prior to MSW.  The beams then emerge from MSW in mixed states and display high-frequency and high-amplitude vacuum oscillations near the detector (at $r=R$).

In each variation, the SDA procedure is challenged to predict \textit{which behavior had occurred} within the supernova envelope: smooth evolution or bipolar oscillations.

To discretize the neutrino path, we record the output of the simulated forward-integration model at  50,001 discretized steps and a step size $dr$ of 0.00004. The optimization procedure uses the same grid.  The units for distance ($r = [0:2]$) are arbitrary, in keeping with previous publications~\cite{armstrong2017optimization,armstrong2020inference}.  These numbers ensure that bipolar oscillation frequency would be well resolved.  Measurements of $P_{1,z}$ and $P_{1,z}$ are taken at the final location ($n=50,001$) and at two locations within 1,500 steps of that final location.  These locations lie sufficiently far beyond the MSW resonance that the vacuum term dominated the Hamiltonian there.  The two additional locations are varied, to determine the solution's robustness to the specific choices of locations.  In total, we conduct 67 independent experiments, corresponding to 67 distinct choices for the locations of the second and third measurement locations. For each these 67 experiments, four paths are searched, beginning at randomly generated initial conditions for state variables.  

The forward integration is performed by Python's odeINT package, which discretizes via an adaptive step size.  The optimization is performed by the open-source Interior-point Optimizer (Ipopt)~\cite{wachter2009short}, which employs a Hermite-Simpson method of discretization and a constant step size.  The discretization of state space, calculations of the model Jacobean and Hessian matrices, and the annealing procedure are performed via an interface with Ipopt that was written in C and Python~\cite{minAone}.  Simulations are run on a computing cluster equipped with 201 GB of RAM and 24 GenuineIntel CPUs (64 bits), each with 12 cores. 

\section{Result} \label{sec:results}

Key results are as follows:
\begin{itemize}
  \item Sampling the $P_{z}$ components of the neutrino beams at multiple vacuum-regime locations reliably predicted \emph{whether} synchronous evolution or bipolar oscillations occurred at earlier radii. 
  \item For the case of bipolar oscillations, and given three measurement locations, the degree to which the measurements sampled the vacuum oscillation amplitude correlated strongly with the strength of the predicted amplitude of earlier bipolar oscillations.  Further, if the vacuum oscillation amplitude was well sampled, the Fourier transforms of the evolution of the $P_z$ components at early radii captured some degree of the complexity of the true waveform.
  \item Using two, rather than three, radial locations, the procedure correctly inferred that bipolar oscillations had occurred, but poorly predicted their amplitude.
  \item A preliminary examination suggests that performing parameter estimation in addition to state estimation, using multiple measurements in vacuum, will be significantly more challenging than performing state estimation alone.
\end{itemize}

\subsection{Prediction of synchronous evolution} \label{sec:resultsSynch}

\begin{figure*}[htb] 
  \includegraphics[width=0.49\textwidth]{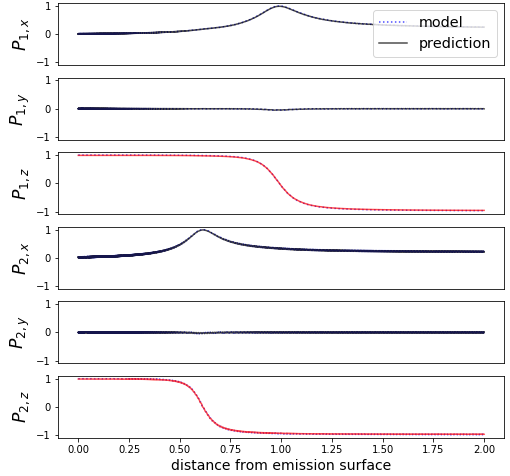}
  \includegraphics[width=0.49\textwidth]{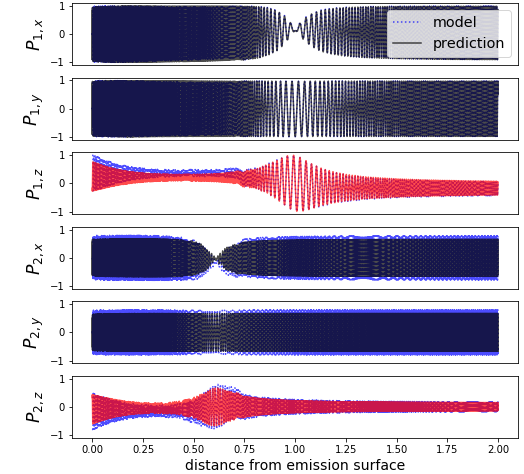}
  \caption{\textbf{True and predicted flavor evolution histories given three measurements of $P_{1,z}$ and $P_{2,z}$ near the detector.} From Top, the columns are: $P_{x}$, $P_{y}$, $P_{z}$ for Beam 1, and $P_{x}$, $P_{y}$, $P_{z}$ for Beam 2.  Black and red lines are predictions for unmeasured and measured state variables, respectively; true model evolution is dotted blue.  \textit{Left}: The initial conditions -- unknown to the SDA procedure -- for $P_{1,z}$ and $P_{2,z}$ were, respectively: +1.0, +1.0, yielding smooth synchronous evolution through the MSW resonance.  \textit{Right}: The initial conditions were instead +1.0 and -0.8, yielding bipolar oscillations.  The units for distance ($r=[0:2]$) are arbitrary, in keeping with previous publications (Refs.~\cite{armstrong2020inference,armstrong2017optimization}).}
 \label{fig2}
\end{figure*}

The left panel of Figure~\ref{fig2} shows the true (dotted blue) versus predicted (solid) state variable evolution for the case in which initial conditions on $P_{1,z}$ and $P_{2,z}$ were +1.0 and +1.0: pure electron-flavor states.  The measured and unmeasured states are in solid red and black, respectively.  These initial conditions yield smooth, synchronous evolution through the MSW resonance, as described in Section~\ref{sec:model}.  In this case, measurements of $P_{1,z}$ and $P_{2,z}$ were taken at three locations in vacuum.  

Here we remind the reader that the measurements used to obtain this prediction were the measurable state variables $P_{1,z}$ and $P_{2,z}$ at three locations near the detector, outside the matter-dominated region; that is: at three out of the 50,001 discretized locations on the path.  Given this sparse information -- which captured no vacuum oscillations, the procedure correctly inferred that the beams had been emitted in aligned pure states and that no bipolar oscillations had  occurred.  This result was robust to ten percent noise added to the measurements of $P_{1,z}$ and $P_{2,z}$ (not shown).

\subsection{Prediction of bipolar oscillations} \label{sec:resultsBipolar}

The right panel of Figure~\ref{fig2} shows the true (dotted blue) versus predicted (solid) state variable evolution for one of the 67 experiments with initial conditions on $P_{1,z}$ and $P_{2,z}$ set to +1.0 and -0.8, respectively.  In the simulation obtained by forward integration, nearly oppositely-aligned, the beams' interactions yield bipolar oscillations.  Given three measurements, which collectively were able to sample the amplitude of vacuum oscillations near Earth, the procedure predicted that bipolar oscillations had occurred at earlier radii.  The prediction of the frequency of these bipolar oscillations was robust to ten percent noise added to the measurements of $P_{1,z}$ and $P_{2,z}$ (not shown).

We sought to quantify in more detail the degree to which the structure of bipolar oscillations at early radii was predicted via observations at later radii in vacuum, over all 67 experiments (as noted in Section~\ref{sec:expersDetails}, the 67 experiments represent 67 distinct choices of two out of three measurement locations: those two lying within 1500 discretized steps of the final location at $n=50,001$).

\begin{figure*}[htb] 
  \includegraphics[width=\textwidth]{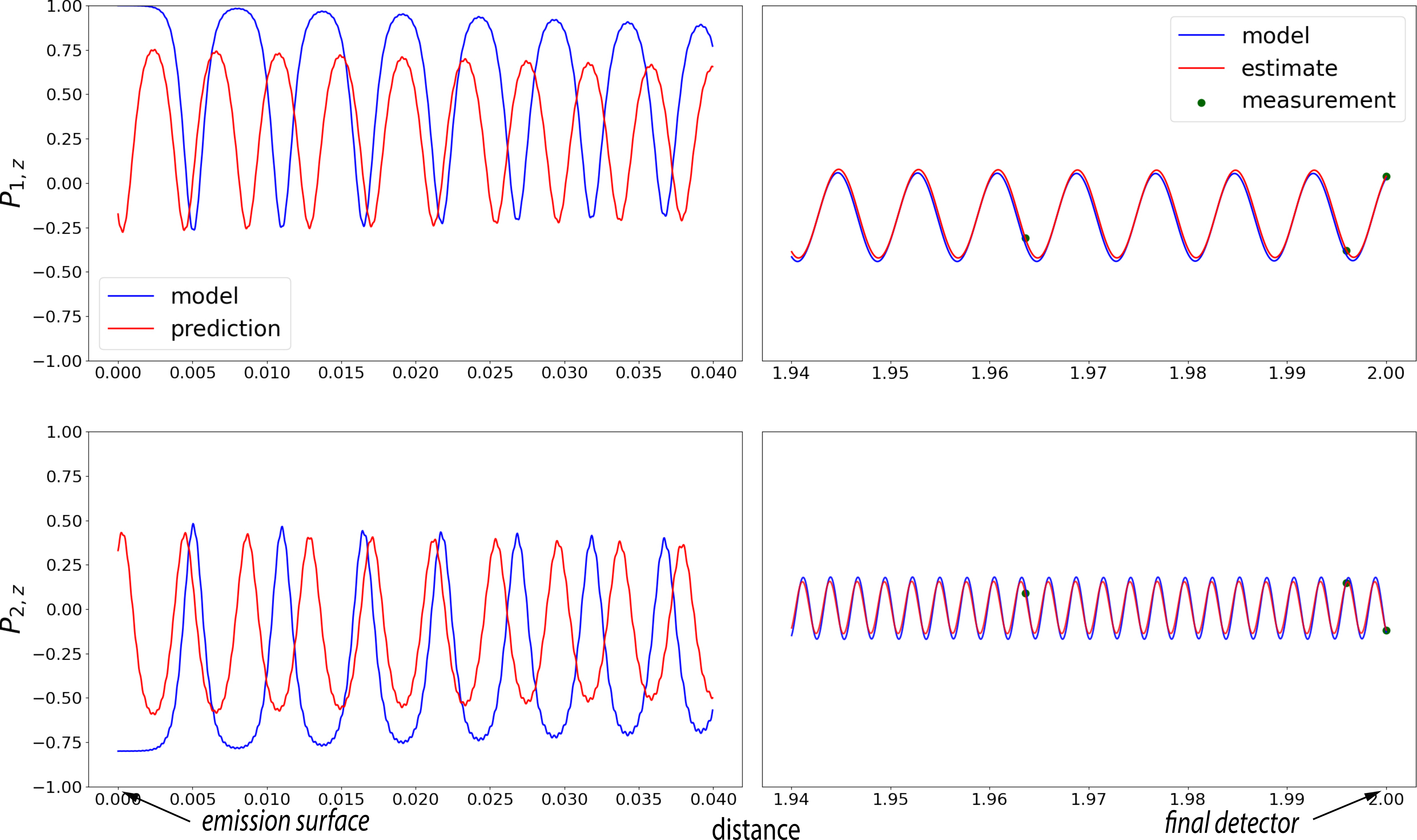}
  \caption{\textbf{Estimate of flavor oscillations in vacuum, juxtaposed with prediction of bipolar oscillations at earlier radii: a "good" example.}  \textit{Top and bottom}: $P_{1,z}$ and $P_{2,z}$, respectively.  \textit{Right}: estimate during observations window in vacuum, with final detector location at far right.  Observation locations denoted by green dots; blue and red are true evolution versus estimate, respectively.  \textit{Left}: prediction of bipolar oscillations at earlier radii.  Note that adequate sampling of the vacuum oscillations amplitude (right) yields a strong prediction of the bipolar oscillations amplitude at earlier radii (left).}
 \label{fig3}
\end{figure*}

First we offer one "good" and one "bad" representative example, out of the 67 total.  Figure~\ref{fig3} shows the "good." The right panels in Figure~\ref{fig3} show the flavor evolution in vacuum near the detector ($n=[48500:50001]$) for $P_{1,z}$ (top) and $P_{2,z}$ (bottom).  True versus predicted are blue and red, respectively; green circles denote the observation locations.  The left panels show the corresponding predicted earlier bipolar oscillations ($n=[0:1000]$). 

Note that together, the three observations (Figure~\ref{fig3} right) capture well the amplitude of vacuum oscillations --- and that the corresponding prediction of earlier bipolar oscillation amplitude is strong (Figure~\ref{fig3} left)\footnote{The difference between true versus predicted initial conditions at $r=0$ is likely due to different discretization methods employed by the optimization-versus-integration procedures.}.  

To further analyze the structure of the predicted bipolar oscillations -- that is, for $n=[0:1000]$, we examined the Fourier decomposition of the evolution of the two $P_z$ components in that region.  A Fourier decomposition was called for because those bipolar oscillations may evolve in radius within that region, and may not be strictly sinusoidal.  

\begin{figure}[htb] 
  \includegraphics[width=0.49\textwidth]{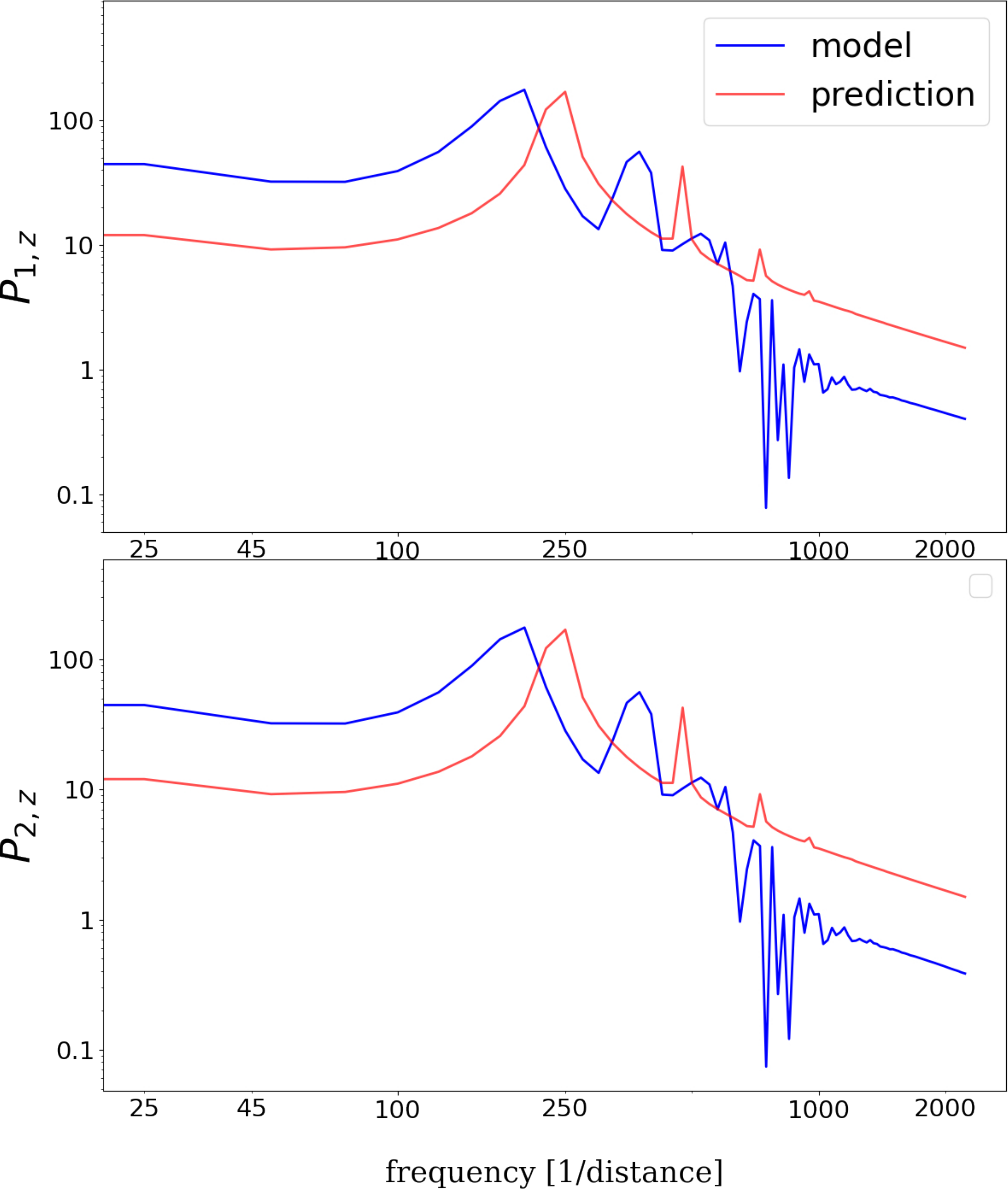}
  \caption{\textbf{The Fourier decomposition of the region of bipolar oscillations shown at left in Figure~\ref{fig3}}, for $P_{1,z}$ (top) and $P_{2,z}$ (bottom).  Blue and red are true and predicted, respectively.  The predicted amplitude of the first harmonic is well matched to the true value, and some complex structure is captured.}
 \label{fig4}
\end{figure}

Figure~\ref{fig4} shows the resulting Fourier power spectrum, for $P_{1,z}$ (top) and $P_{2,z}$ (bottom), where blue and red are true versus predicted, respectively.  The predicted amplitude of the strongest harmonic is well matched to the true value, as was indicated by the wave-forms at left in Figure~\ref{fig3}\footnote{The difference in the precise value of the predicted versus true peak frequency corresponds to the minimum difference set by the sampling rate.  That nonzero difference is likely due to different discretization methods employed by the optimization-versus-integration procedures.}.

Further, and more interestingly, the predicted Fourier transform captures to some degree the complexity of structure present in the correct solution.  First, the peak frequency is not a delta function, but rather has a finite width, indicating that its value is evolving within the range of $n=[0:1000]$.  Second, the second harmonic is also predicted. (The features at higher frequencies are likely due to differences in the discretization methods used by the Python forward integration versus the optimizer.)

\begin{figure*}[htb] 
  \includegraphics[width=\textwidth]{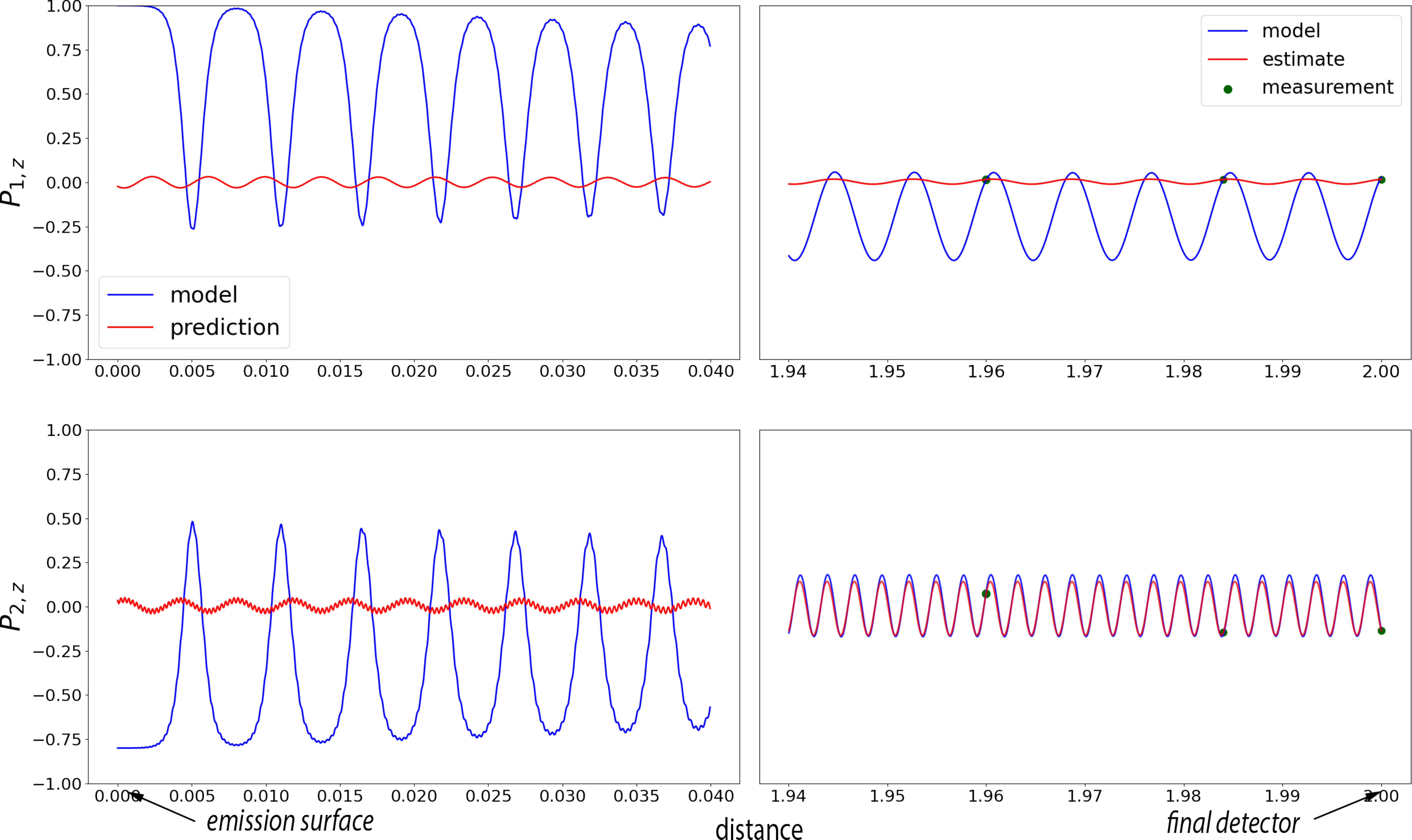}
  \caption{\textbf{Estimate of flavor oscillations in vacuum, juxtaposed with prediction of bipolar oscillations at earlier radii: a "bad" example.}  \textit{Top and bottom}: $P_{1,z}$ and $P_{2,z}$, respectively.  \textit{Right}: estimate during observations window in vacuum, with final detector location at far right.  Observation locations denoted by green dots; blue and red are true evolution versus estimate, respectively.  \textit{Left}: prediction of bipolar oscillations at earlier radii.  The poor estimate of vacuum oscillation amplitude (right) yields a poor prediction of bipolar oscillation amplitude earlier (left).}
 \label{fig5}
\end{figure*}

\begin{figure}[htb] 
  \includegraphics[width=0.49\textwidth]{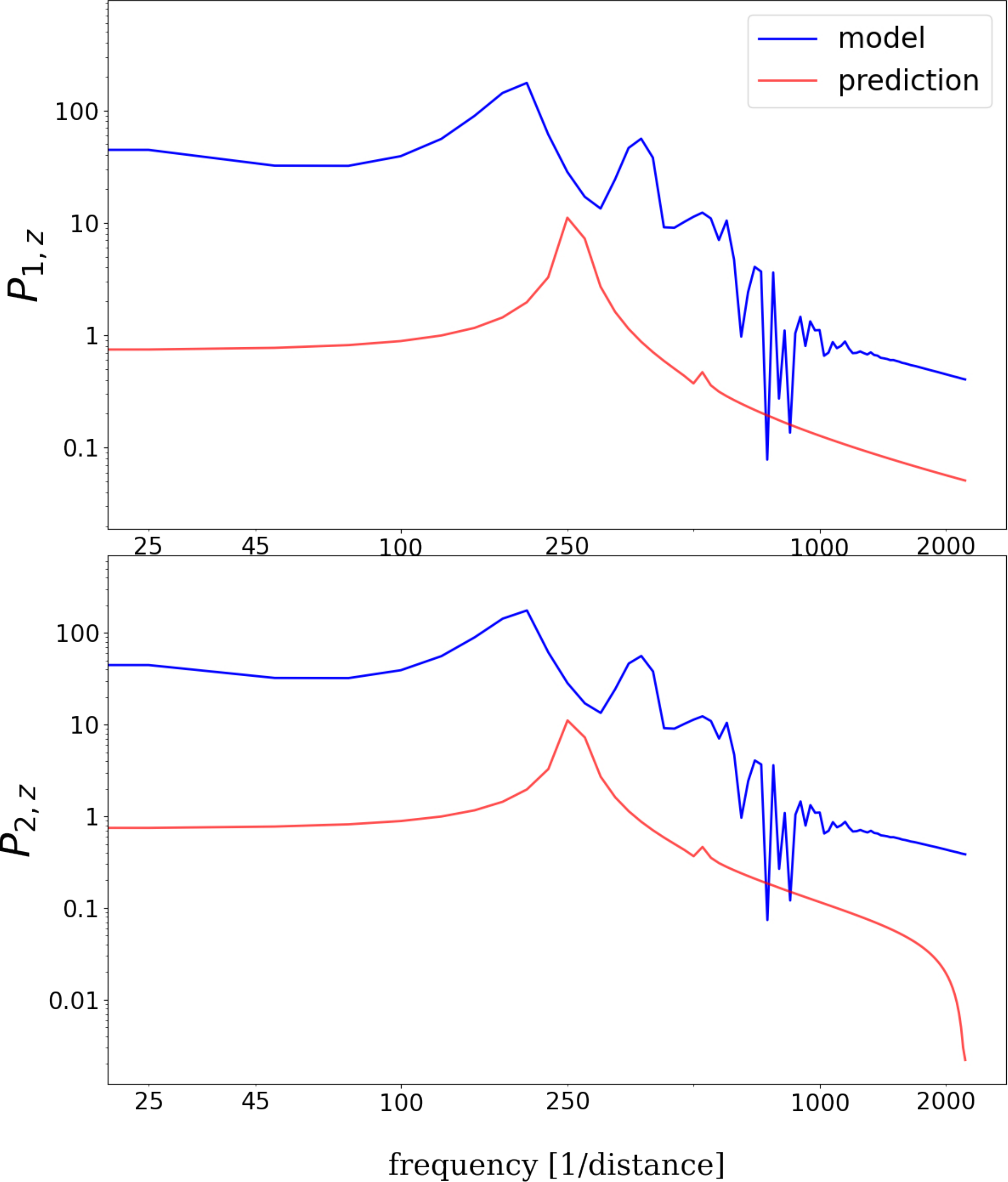}
  \caption{\textbf{The Fourier decomposition of the region of bipolar oscillations shown at left in Figure~\ref{fig5}}.  The predicted amplitude of the first harmonic is poor, and complex structure is not captured well.}
 \label{fig6}
\end{figure}

Figure~\ref{fig5} shows a representative "bad" solution out of the 67 total; it is formatted identically to Figure~\ref{fig3}.  Note that the three measurements (right) poorly sample the vacuum amplitude, and that that poor estimate is reflected in a poor prediction of the bipolar oscillation amplitude (Figure~\ref{fig5} left, top and middle).  The Fourier transforms of Figure~\ref{fig6}, akin to Figure~\ref{fig4} for the "good" example, also reflect a poorer match to the power in the first harmonic of the bipolar oscillations waveform, and the complexity of the Fourier decomposition of the true solution is not captured strongly. 

These two examples alone suggest a pattern: in the experiments where the multiple measurements in the detector region are able to sample the full extent of the vacuum oscillation amplitude, the prediction of the bipolar oscillation waveform near the source becomes significantly better. To quantify the \lq\lq goodness\rq\rq\ of the bipolar oscillation prediction, the metric that we used was the absolute value of the difference between the peak amplitudes of the strongest frequency in the Fourier transforms (FFTs), of the true and predicted $P_z$ waveforms near the source (the first 1000 grid points), summed over both the neutrino energies.  We will call this quantity $\Delta \text{FFT}^{\textit{\,peak, bipolar}}$ and formally write:  
\begin{widetext}
\begin{equation}
    \Delta \text{FFT}^\textit{\,peak, bipolar} = \sum_j \left| \text{max}\left\{ \text{FFT}\left[P_{j,z}^\text{(true)}(n \in [0:1000])\right] \right\} - \text{max} \left\{ \text{FFT}\left[P_{j,z}^\text{(pred)}(n \in [0:1000])\right] \right\} \right|,
    \label{eq:deltafftpeak}
\end{equation}
\end{widetext}
where \lq\lq max\rq\rq\ refers to the maximum strength of the FFT across all Fourier modes (except the zero-frequency mode).  We correlated that quantity with the degree to which the observations captured the vacuum oscillation amplitudes of both neutrinos.  Formally, we defined the difference in true versus estimated amplitude near the detector as $\Delta P_{z}^\textit{\,ampl, vacuum}$:
\begin{widetext}
\begin{equation}
    \Delta P_{z}^\textit{\,ampl, vacuum} = \sum_j \left| \text{ampl}\left\{ P_{j,z}^\text{(true)}(n \in [48500:50000]) \right\} - \text{ampl}\left\{ P_{j,z}^\text{(pred)}(n \in [48500:50000]) \right\} \right|,
    \label{eq:pzampldiff}
\end{equation}
\end{widetext}
where \lq\lq ampl\rq\rq\ refers to the amplitude of each $P_z$ waveform over the specified domain (in this case, the discrete grid locations numbered 48500 to 50000).

Indeed, across all 67 experiments, we discovered that these two metrics were well correlated. This is shown in Fig.~\ref{fig:FFTvsPzampl}, where $\Delta \text{FFT}^\textit{\,peak, bipolar}$ ($y$-axis) is plotted against $\Delta P_{z}^\textit{\,ampl, vacuum}$ ($x$-axis).

\subsection{Varying the number of measurements} \label{sec:resultsVaryMeas}

Prior to conducting the experiments described above, which employed three measurement locations, we had attempted to find solutions using just one measurement location for the $P_z$ components, and then using two locations.  Results were as follows.

Measuring the $P_z$ components only at the final location (at $r=R$) yielded zero model dynamics (not shown).  Adding a second measurement location yielded a correct inference that bipolar oscillations had occurred, although the amplitude of those oscillations was predicted poorly (not shown).  Adding a third measurement significantly improved the prediction of the amplitude of bipolar oscillations, as shown in Figure~\ref{fig2} Right Panel and Figure~\ref{fig3}.

To interpret this result, one must recognize that measuring $P_z$ at more than one location yields information about the \textit{derivative} of $P_z$ -- which depends on the unmeasureable variables $P_x$ and $P_y$.  As one increases the number of locations at which $P_z$ is measured, one is effectively reconstructing the dynamics of the imaginary $P_x$ and $P_y$.  See Section~\ref{sec:disc}.

\subsection{Prediction with an unknown parameter in the model's matter potential} \label{sec:resultsParamEst}

With a nonlinear model, rendering a single parameter to be an unknown quantity significantly increases the mathematical challenge for the SDA procedure, compared to state estimation alone.  We sought to ascertain how well the SDA design described in this paper would navigate such an increase in complexity.  To this end, we repeated the experiments, this time setting one model parameter to be an unknown quantity to be estimated along with the state variables.  We chose as this unknown parameter the constant coefficient $C_m$ in the matter potential (as described in \textit{Model}), because the matter potential is of keen theoretical interest and may impart a signature upon a detection.  

For both the cases with and without bipolar oscillations, we initialized ten independent paths, using as measurements the $P_{1,z}$ and $P_{2,z}$ values at three locations near $r=R$, as before.  The true value of $C_m$ was 3308.0, and the permitted search range was: [0:10,000.]

For both cases with and without bipolar oscillations, the estimates of $C_m$ were scattered within the permitted search range, with half of the paths estimating the upper bound of 10,000.  For each estimate, we confirmed via forward integration that the corresponding state variable evolution was as it should be, were the true value of $C_m$ indeed the estimated value.  Importantly, for both cases, the SDA procedure still correctly inferred \textit{whether} bipolar oscillations occurred, and captured the amplitude and frequency of those oscillations as faithfully as it had in the original experiment that had taken $C_m$ to be known.  We conclude that the three measurements of $P_z$ near $r=R$ contain significantly more information about the state variable evolution than they do about the specific strength of the matter potential.  Adding more realistic complexity to the matter potential, for example, including shocks, may improve the procedure's ability to home in on its precise form.  Moreover, a detailed study of the procedure's ability to handle parameter estimation will require an examination of the model's sensitivity to specific parameter values.

\section{Discussion} \label{sec:disc}

We have made significant progress beyond previous work, having eliminated assumptions regarding flavor content throughout the matter-dominated regime, and instead inferring flavor evolution histories via measurements at accessible locations in vacuum.  We have learned that obtaining a measurement of polarization vector component $P_z$ at multiple locations in vacuum yields information about whether bipolar oscillations occurred at earlier radii, prior to the MSW transition.  

\subsection{How do multiple measurements of $P_z$ in vacuum predict bipolar oscillations at emission?}

As noted in Section~\ref{sec:resultsBipolar}, one measurement in vacuum of the $P_z$ components of the two neutrino beams yielded failed inference of the flavor evolution history.  Two measurement locations correctly showed whether bipolar oscillations had occurred at emission, but with poor matches to the oscillation amplitude.  Three measurement locations significantly enhanced that amplitude prediction.  What is the significance of "at least three measurements"?

This question brings to mind the time-delay embedding theorem from dynamical systems.  At the core of that prescription is the notion that one can represent a state space in $n$ variables, or equivalently in one variable at $n$ distinct temporal locations~\cite{ruelle1979ergodic}.  The concept is intimately related to the information contained in the \textit{derivatives} of a time series~\cite{eckmann1985ergodic}.  The relevant scenario for our purposes is that, taken together, multiple measurements of $P_z$ represent the derivative of $P_z$.  According to Equation~\ref{eq:model}, that derivative is dictated in part by the instantaneous values of $P_x$ and $P_y$.  That is: the derivative of $P_z$ contains information about $P_x$ and $P_y$ -- and hence \textit{phase information.}


A single measurement location of $P_{1,z}$ and $P_{2,z}$ contains no information regarding the relative phases of the respective polarization vectors.  Two measurement locations, however, yield some crude approximation of the derivatives, and hence can reconstruct to some degree the instantaneous values of $P_x$ and $P_y$.  Thus, a pair of measurements of $P_z$ contains some information about the relative phases of the two beams, and hence \textit{whether} bipolar oscillations could have occurred at prior radii.

\begin{figure}[H] 
  \includegraphics[width=0.49\textwidth]{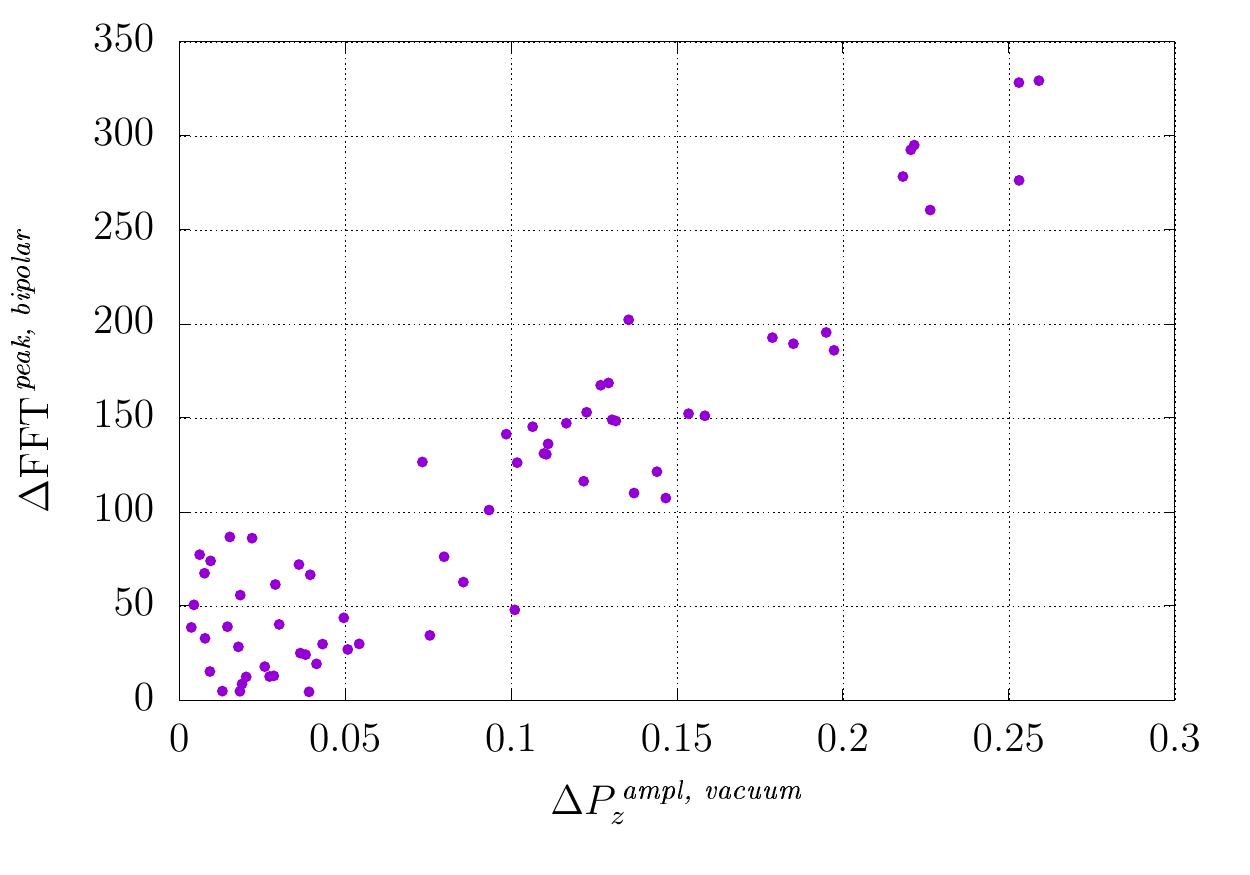}
  \caption{Correlation between the goodness of the bipolar oscillation prediction, as quantified by the difference between the peak FFT amplitudes of the true and predicted $P_z$ waveforms in the source region ($\Delta \text{FFT}^\textit{\,peak, bipolar}$, as defined in Eq.~\eqref{eq:deltafftpeak}), and the difference between the true and predicted $P_z$ oscillation amplitudes in the detector region ($\Delta P_{z}^\textit{ampl, vacuum}$, as defined in Eq.~\eqref{eq:pzampldiff}). Each dot on the plot represents one experiment.}
 \label{fig:FFTvsPzampl}
\end{figure}

Adding yet a third measurement further improves the accuracy of the predicted derivative of the $P_z$ components.  Having a third measurement greatly increases the likelihood of sampling the amplitude of the vacuum oscillation waveforms of the individual neutrino modes in the detector region. As shown in Fig.~\ref{fig:FFTvsPzampl}, sampling the full amplitude of the vacuum oscillations is well-correlated with soundly predicting the amplitude of bipolar oscillations in the source region. Based on this correlation, we might expect that, increasing the number of measurements beyond three would further improve the bipolar oscillation predictions near the source. To this end, we conducted two preliminary tests that employed four and five measurement locations, respectively, using just one set of measurement locations for each test.  Both yielded excellent predictions, comparable to the best result obtained over all 67 experiments that had employed three measurements --- i.e. Figure~\ref{fig2}, Right Panel.  This finding is unsurprising: the more independent locations of $P_z$ sampled in vacuum, the more precisely its derivative can be estimated.

Finally, the reader might have noted that, in measuring the $P_z$ vector components at three distinct locations, we replaced the six boundary conditions used by forward integration by six different boundary conditions.  In the case of forward integration, the six are: all measurable ($P_z$) and unmeasurable state variables ($P_x$ and $P_y$) for both beams at a single physically inaccessible location ($r=0$).  By contrast, within the SDA formulation, the six boundary conditions were the \textit{measurable} $P_z$ for both beams at three physically accessible regions (in vacuum near $r=R$).  In the future, it might prove instructive to formalize a translation between these two formalisms.


\section{ACKNOWLEDGEMENTS}

E.~A., A.~A., M.~S., and M.~I. acknowledge an Institutional Support for Research and Creativity grant from New York Institute of Technology.  E.~A. acknowledges NSF grant 2139004.  S.~M. acknowledges the NSF summer Research Experience for Undergraduates program. The work of A.~V.~P. was supported by the U.S. Department of Energy under contract number DE-AC02-76SF00515..  As always, eternal thanks to the good people of Doylestown, Ohio.

\bibliography{bib,bib_eve_collisions,bib_armstrong2020paper}
\end{document}